\documentclass[aps,prd,nofootinbib,showkeys,floatfix,twocolumn]{revtex4-1}
\usepackage{amsmath}
\usepackage{amssymb}
\usepackage{graphicx}
\usepackage{xspace}
\usepackage{bbm} 
\usepackage{color}
\usepackage[utf8]{inputenc}
\usepackage{cancel}
\usepackage{slashed}
\usepackage{soul}

\definecolor{mkgreen}{rgb}{0.2,.70,.3}

\definecolor{tobycolour}{rgb}{.5,.0,.5}

\newcommand{\FSadd}[1]{{\textcolor{blue}{#1}}}

\newcommand{\eVdist}{\kern-0.06em}

\newcommand\SARAH{{\tt SARAH}\xspace}
\newcommand\Vevacious{{\tt Vevacious}\xspace}
\newcommand\SPheno{{\tt SPheno}\xspace}

\newcommand{\be}{\begin{equation}}
\newcommand{\ee}{\end{equation}}
\newcommand{\bea}{\begin{eqnarray}}
\newcommand{\eea}{\end{eqnarray}}

\begin{document}
\vspace{1cm}

\title{Spontaneous Charge Breaking in the NMSSM -- Dangerous or not?}

\hfill \parbox{5cm}{\vspace{ -1cm } \flushright BONN-TH-2017-02 \quad KA-TP-07-2017}

 \author{Manuel E. Krauss}
 \email{mkrauss@th.physik.uni-bonn.de}

 \author{Toby Opferkuch}
 \email{toby@th.physik.uni-bonn.de}
\affiliation{$^1$Bethe Center for Theoretical Physics \& Physikalisches Institut der 
Universit\"at Bonn, \\
Nu{\ss}allee 12, 53115 Bonn, Germany}

 \author{Florian Staub}
 \email{florian.staub@kit.edu}
 \affiliation{Institute for Theoretical Physics (ITP), Karlsruhe Institute of Technology, Engesserstra{\ss}e 7, D-76128 Karlsruhe, Germany}
 \affiliation{Institute for Nuclear Physics (IKP), Karlsruhe Institute of Technology, Hermann-von-Helmholtz-Platz 1, D-76344 Eggenstein-Leopoldshafen, Germany}

\begin{abstract}
We investigate the impact of charge-breaking minima on the vacuum stability of the NMSSM. We find that, in contrast to Two-Higgs-Doublet Models like the MSSM, at both tree- and loop-level there exists global charge-breaking minima. Consequently, many regions of parameter space are rendered metastable, which otherwise would have been considered stable if these charge-breaking minima were neglected. However, the inclusion of these new scalar field directions has little impact on otherwise metastable vacuum configurations.
\end{abstract}

\maketitle

\section{Introduction} 
At first glance, the discovery of a standard model (SM)-like Higgs boson with a mass of approximately 125~GeV \cite{Aad:2012tfa,Chatrchyan:2012xdj} appears to be a huge success of supersymmetry (SUSY) and in particular of the minimal supersymmetric standard model (MSSM). In contrast to other ideas to extend the SM, SUSY predicts that the Higgs boson shouldn't be significantly heavier than the $Z$-boson if new physics is around the TeV scale, see e.g. Ref.~\cite{Martin:1997ns} and references therein. Other avenues such as technicolor  prefer the natural mass range for the Higgs to lie at scales well above the measured mass. On the other hand,  closer investigation shows that the situation is more complicated in the MSSM as the Higgs mass requires large radiative corrections to be compatible with experimental data. The main source of these corrections are the superpartners of the top, the stops. In order to maximise their contributions to the Higgs mass, one needs to consider scenarios in which they are maximally mixed \cite{Brignole:1992uf,Chankowski:1991md,Dabelstein:1994hb,Pierce:1996zz}. This can be dangerous because it can lead to the presence of charge- and colour-breaking vacua whereby the stops receive vacuum expectation values (VEVs) \cite{Camargo-Molina:2013sta,Blinov:2013fta,Chowdhury:2013dka,Camargo-Molina:2014pwa}. Since the tunnelling rate to these vacua is typically large, this results in tension between an acceptable Higgs mass and a sufficiently long-lived electroweak (EW) breaking vacuum. Consequently, SUSY models which can enhance the Higgs mass at tree-level are especially appealing. The simplest such extension is to add a scalar singlet, resulting in the next-to-minimal supersymmetric standard model (NMSSM), yields $F$-term contributions, which raise the tree-level Higgs mass \cite{Ellwanger:2009dp,Ellwanger:2006rm}. This significantly reduces the need for large loop corrections. As a result, large stop mixing is no longer necessary. Therefore, the vacuum stability problems of the MSSM are cured as well as reducing the EW fine-tuning \cite{BasteroGil:2000bw,Dermisek:2005gg,Dermisek:2006py,Dermisek:2007yt,Ellwanger:2011mu,Ross:2011xv,Ross:2012nr,Kaminska:2013mya}. However, the extended Higgs sector in the NMSSM introduces new couplings which can potentially destabilize the EW vacuum. The vacuum stability in the NMSSM has been studied in the past at tree-level \cite{Ellwanger:1996gw, Ellwanger:1999bv, Kanehata:2011ei, Kobayashi:2012xv,Agashe:2012zq}, and also with one-loop corrections \cite{Beuria:2016cdk}. Potentially dangerous parameter ranges have been identified in these works. However, all these studies made the assumption that charge is conserved at the global minimum of the scalar potential, i.e. the charged Higgs boson VEVs were neglected. This was motivated to some extent as it has been shown that the global minimum of two-Higgs-doublet models is always charge conserving at tree-level \cite{Barroso:2005sm}. However, for non-vanishing singlet--doublet interactions this is no longer the case \cite{Muhlleitner:2016mzt} and one must in principle always take these VEVs into account. The aim of this letter is to discuss the impact of  charged Higgs VEVs on the vacuum stability in the NMSSM. We start in Sec.~\ref{sec:analytic} with a discussion of the scalar potential, before we show the numerical results in Sec.~\ref{sec:numerics}. We conclude in Sec.~\ref{sec:conclusion}.

\section{Spontaneous charge breaking in the NMSSM}
\label{sec:analytic}
We consider in the following the NMSSM with a $\mathbb{Z}_3$ to forbid all dimensionful parameters in the superpotential. The superpotential reads
\begin{equation}
W_{\rm NMSSM} = \lambda \hat H_d \hat H_u \hat S + \frac{1}{3}\kappa \hat S^3 + W_Y \,,
\end{equation}
with the standard Yukawa interactions $W_Y$ as in the MSSM. The additional soft-terms in comparison to the MSSM are
\begin{equation}
- \mathcal L_{\rm soft} \supset \left(T_\lambda H_d H_u S + \frac13 T_\kappa S^3 + \text{h.c.} \right) + m_s^2 |S|^2\,,
\end{equation}
where we have used the common parametrisation for the trilinear soft terms
\begin{equation}
 T_\lambda = A_\lambda \lambda \,,\hspace{1cm} T_\kappa = A_\kappa \kappa\,.
\end{equation}
After electroweak symmetry breaking, the scalar singlet $S$ obtains a VEV $v_S$ which generates an effective Higgsino mass term $\mu_{\rm eff}$ 
\begin{equation}
\mu_{\rm eff} = \frac{1}{\sqrt{2}} \lambda v_S \,.
\end{equation}
Using the three minimisation conditions of the potential, the Higgs sector in the NMSSM is specified at tree-level by six parameters:
\begin{eqnarray}
\lambda\,, \quad \kappa\,, \quad A_\lambda\,, \quad A_\kappa\,, \quad \mu_{\rm eff}\,, \quad  \tan\beta\,,
\end{eqnarray}
with the ratio $\tan\beta=\frac{v_u}{v_d}$ of the doublet VEVs. \\
However, we have so far neglected the possibility that charged Higgs bosons can acquire VEVs. In order to include this possibility, one needs 
to check for the global minimum of the scalar potential resulting from the following VEVs:
\begin{align}
\begin{pmatrix}
\langle H_d^0 \rangle \\
\langle H_d^- \rangle
\end{pmatrix} &= \frac{1}{\sqrt{2}}
\begin{pmatrix}
v_d \\
v_m
\end{pmatrix}\,, \quad
\begin{pmatrix}
\langle H_u^+ \rangle \\
\langle H_u^0 \rangle
\end{pmatrix} = \frac{1}{\sqrt{2}}
\begin{pmatrix}
v_p \\
v_u
\end{pmatrix} \\
&\qquad\langle S \rangle = \frac{v_S}{\sqrt{2}}
\end{align}
One can reduce this five-dimensional problem via an $SU(2)$ gauge transformation to eliminate one of the charged Higgs VEVs. This turns out to be more robust for the numerical evaluation, 
but for the current discussion we keep the more intuitive form with all five VEVs. \\
The scalar potential of the Higgs sector in the NMSSM with these five VEVs consists of $F$-,  $D$- and soft-terms
\begin{align}
V_{\rm Full} = V_F + V_D + V_{\rm soft} \,,
\end{align}
with
\begin{align}
V_F = & \frac{1}{4} \Big(\lambda  v_S^2 \left(\lambda  \left(v_d^2+v_m^2+v_p^2+v_u^2\right)+2 \kappa  (v_m v_p-v_d v_u)\right)\nonumber \\
&+ \lambda ^2 (v_m v_p-v_d  v_u)^2+\kappa ^2 v_S^4\Big)\,, \\
V_D = & \frac{1}{32} \Big(g_1^2 \left(v_d^2+v_m^2-v_p^2-v_u^2\right)^2 \nonumber \\
&+ g_2^2 \Big(v_d^4+v_m^4 +2 v_d^2 \left(v_m^2+v_p^2-v_u^2\right)+8 v_d v_m v_p
v_u\nonumber \\
& -2 v_m^2 \left(v_p^2-v_u^2\right)+\left(v_p^2+v_u^2\right)^2\Big)\Big)\,, \\
V_{\rm Soft} = &\frac{1}{2} \Big(m_{H_d}^2 \left(v_d^2+v_m^2\right)+m_{H_u}^2 \left(v_p^2+v_u^2\right) +m_S^2 v_S^2\Big)  \nonumber \\
&+ \frac{v_S}{6} \left(\sqrt{2} T_\kappa v_S^2+3 \sqrt{2} T_\lambda
   (v_m v_p-v_d v_u)\right)\,.
\end{align}
Before we continue, we can check if parameter points exist, for which the global minimum of the potential is charge breaking. In order to do so, we compute
\begin{equation}
\Delta V = V_{\rm Full} - V_{\rm Full}\big|_{v_m=v_p=0}\,.
\end{equation}
Together with the relation between $A_\lambda$ and the charged Higgs mass $m_{H^+}$ 
\begin{equation}
A_\lambda = \frac{\lambda  t_\beta \left(4 m_{H^+}-v^2 \left(g_2^2-2   \lambda ^2\right)\right)-4 \kappa \mu_{\rm eff}^2 \left(t^2_\beta+1\right)}{4   \lambda  \mu_{\rm eff} \left(t^2_\beta+1\right)}\,,
\end{equation}
where $t_\beta = \tan\beta$, we get  in the limit $t_\beta\to 1$, $v_m \to 0$ \footnote{Don't be confused about the presence of $v_d$,$v_u$, $v_S$ and $v$, $\mu_{\rm eff}$ at the same time:
$v_d$,$v_u$, $v_S$ are free degrees of freedom of the scalar potential, while $v$ and $\mu_{\rm eff}$ correspond to the necessary values to get a local minimum 
with correct EW symmetry breaking. Again, the choice $v_m\to0$ can always be made using a $SU(2)$ gauge transformation.}
\begin{align}
\Delta V &= \frac{1}{32} v_p^2 \Big(g_2^2 \left(2 v_d^2-2 v^2+v_p^2+2 v_u^2\right)-16 \mu_{\rm eff}^2+8 \lambda ^2
   v_S^2 \nonumber \\
&\qquad+ g_1^2 \left(v_p^2-2 v_d^2+2 v_u^2\right)+8  m_{H^+}\Big)\,.
\end{align}
Thus, one can see that in particular for large $\mu_{\rm eff}$ it is possible to get very deep charge-breaking (CB) minima below those which are charge-conserving (CC). \\

We now seek to gain some additional insight into the behaviour of the potential and, in particular, regions where the CB minima are potentially dangerous. The most promising directions in field space to discover deep minima are those in which 
either the $F$- or $D$-terms vanish. Since we are in general interested in points with sizeable $\lambda$ couplings in order to get a large enhancement for the Higgs mass, the most stabilising effect of the potential can be expected to come from the $F$-terms. It is actually not possible to find any $F$-flat directions which are charge conserving. However, in the charge-breaking case the $F$-terms vanish for
\begin{equation}
\label{eq:dir}
v_m = v_u\,,\quad v_p = v_d\,,\quad v_S =0 \,.
\end{equation}
In this direction in VEV space the value of the potential is
\begin{equation}
V = \frac18 (v_d^2 + v_u^2) (4 m_{H_d}^2 + 4 m_{H_u}^2 + g_2^2 (v_d^2 + v_u^2)) \,,
\end{equation}
which can be related in the limit $\tan\beta \to 1$ to the input parameters 
\begin{align}
\label{eq:CBdirection}
V&= \frac{1}{8}\left(v_d^2+v_u^2\right) \bigg(8 A_\lambda  \mu_{\rm eff}+ g_2^2  \left(v_d^2+v_u^2\right) \nonumber \\
&\quad+8 \mu_{\rm eff}^2 \left(\frac{\kappa}{\lambda} -1\right)-2 \lambda^2 v^2\bigg)\,.
\end{align}	
From this expression one sees that the following conditions characterise the potentially dangerous regions in which CB minima might develop: ($i$) large $|\lambda|$ and $|\mu_{\rm eff}|$, (ii) either opposite signs for $\lambda$ and $\kappa$ or $|\kappa/\lambda| < 1$ as well as (iii) opposite signs for $A_\lambda$ and $\mu_{\rm eff}$.
Eq.~(\ref{eq:CBdirection}) has to be combined with the condition that all Higgs masses are non-tachyonic at the electroweak vacuum. The condition to have a positive charged Higgs mass
is
\begin{equation}
0 < \frac{1}{4} v^2 \left(g_2^2-2 \lambda ^2\right)+2 \frac{\kappa}{\lambda} \mu_{\rm eff}^2+2 \mu_{\rm eff} A_\lambda \,,
\end{equation}
which for large $\mu_{\rm eff}^2$, prefers $\lambda$ and $\kappa$ of same signs and also either equal signs for  $A_\lambda$ and $\mu_{\rm eff}$ or small $A_\lambda$ compared to $\mu_{\rm eff}$. 
From the positivity condition on the pseudo-scalar masses one can further see that opposite signs for $A_\kappa$ and $\mu_{\rm eff}$ are preferable.
Therefore, combined with Eq.~(\ref{eq:CBdirection}), we see that CB minima are likely to occur if:
\begin{itemize}
\item $|\lambda|$ and $\FSadd{|}\mu_{\rm eff}\FSadd{|}$ are large 
\item $|\kappa/\lambda|<1$ with ${\rm sign}(\kappa)={\rm sign}(\lambda)$
\item $|A_\lambda/\mu_{\rm eff}|<1$
\item sign($A_\kappa$)$=-$sign($\mu_{\rm eff}$)
\end{itemize} 
It is important to note that in these regions, the mostly singlet-like scalar is heavy therefore,
the SM-like Higgs is always the lightest CP-even scalar state. 
\begin{figure}[tb]
\includegraphics[width=\linewidth]{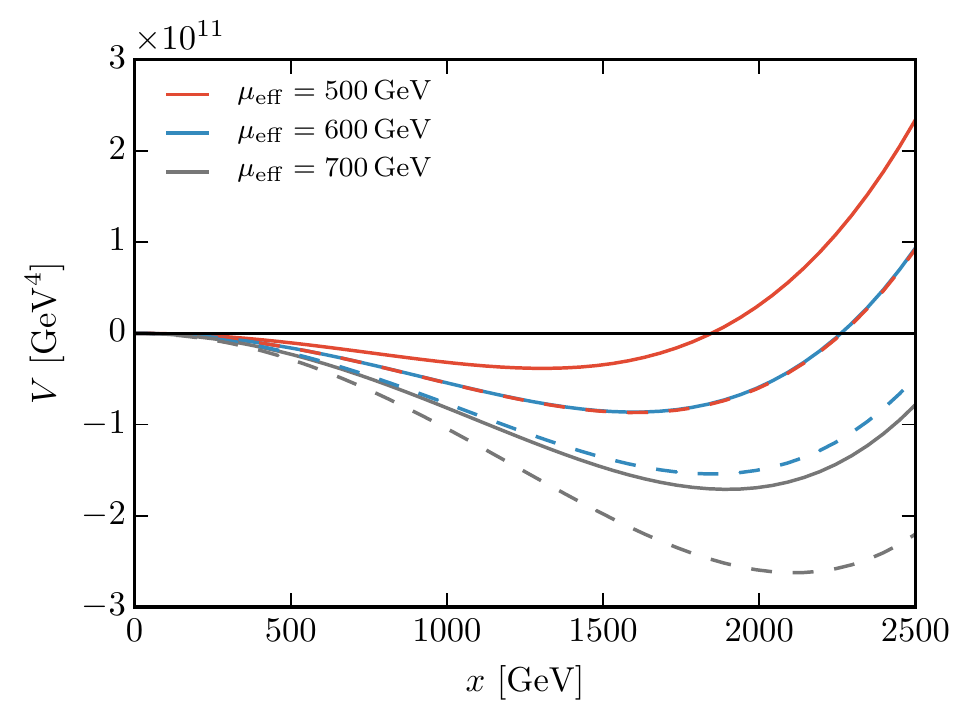} 
\caption{Value of the scalar potential in the direction of vanishing $F$-terms for three different values of $\mu_{\rm eff}$. Here, we have chosen 
 $\kappa=\frac12 \lambda$,  $A_\lambda = 100$ as well as $\lambda=-1$ (full lines) and $\lambda = -2$ (dashed lines).}
\label{fig:analytic}
\end{figure}

\begin{figure}[tb]
\includegraphics[width=\linewidth]{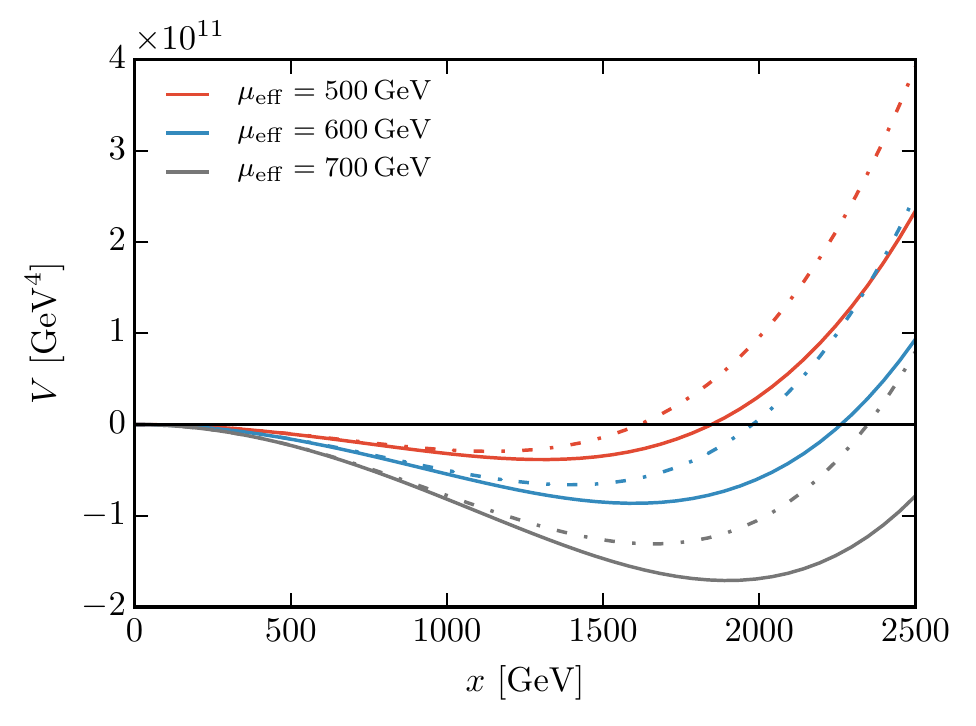} 
\caption{Comparison of the potential in the charge conserving (dash-dotted) and charge-breaking direction  (full) defined by Eqs.~(\ref{eq:dir}) respectively (\ref{eq:dircon}). The same 
parameter choices as in Fig.~\ref{fig:analytic} were made and we show here the case $\lambda = -1$.}
\label{fig:ana_comparison}
\end{figure}

In Fig.~\ref{fig:analytic}, we show the behaviour of the potential in the direction $x=\sqrt{v_d^2+v_u^2+v_m^2+v_p^2}$ for different values of
$\mu_{\rm eff}$. We see in these examples that the minima are in the multi-TeV range and move quickly to larger values with increasing $\mu_{\rm eff}$. Thus, it needs to be checked how efficient the tunnelling to these minima is. In addition, one also needs to compare the tunnelling to these minima with the tunnelling to potential CC minima which don't coincide with the 
electroweak breaking vacuum. One important VEV direction in this context is the one with 
\begin{equation}
\label{eq:dircon}
v_u = v_m = v_p=v_S=0\,,~v_d\neq 0\,,
\end{equation}
in which the potential is given by
\begin{align}
V = & \frac{v_d^2}{2} \bigg(A_\lambda  \mu_{\rm eff}+  \frac{v_d^2}{16} \left(g_1^2+g_2^2\right)\nonumber \\
& \qquad +\mu_{\rm eff}^2 \left(\frac{\kappa}{\lambda} -1 \right)-\frac{1}{4}  \lambda^2 v^2\bigg)\,.
\end{align}
A one-dimensional comparison between the behaviour of the potential in this direction and in the direction defined via Eq.~(\ref{eq:dir}) is shown in Fig.~\ref{fig:ana_comparison}.
As a result, we observe in typical regions of parameter space that CB and CC minima occur at the same time and, both are usually deeper than the correct electroweak vacuum. 
Furthermore, it can be seen from Fig.~\ref{fig:ana_comparison} that the CB minimum is deeper than the non-EW but CC. However, the latter appears at slightly smaller $x$ values.
Consequently, it is not a priori clear to which minima the electroweak state would tunnel to more effectively  -- to the deeper one or the nearer one -- as the field space is highly non-trivial. In these cases, one needs to calculate the tunnelling rate to the different minima in order to be able to judge if the inclusion of charged Higgs VEVs yields additional constraints. 

In general, the decay rate $\Gamma$ per unit volume  for a false vacuum is given in \cite{Coleman:1977py, Callan:1977pt} by
\begin{equation}
 \Gamma / \text{vol.} = A e^{( -B / \hbar)} \left( 1 + \mathcal{O}( \hbar ) \right)\,,
\label{eq:tunneling_time}
\end{equation}
where $A$ is a factor which depends on the eigenvalues of a functional determinant  and $B$ is the bounce action. $A$ is usually taken to be of order the renormalisation 
scale and is less important for the tunnelling rate which is dominated by the exponent $B$. In a multi-dimensional space it only makes sense to calculate $B$ numerically as any approximations, analytic or otherwise, are simply not accurate enough due to the huge sensitivity of $\Gamma$ on $B$. Of course, there are also other directions in VEV space where CB minima might establish. However, an analytical discussion of all these cases doesn't give further insights. We therefore turn directly to the numerical results.

\section{Numerical results}
\label{sec:numerics}
As we have seen so far, one can find new vacua in the NMSSM when including the possibility of spontaneous charge breaking. However, it needs to be clarified how important the study of these minima is. Therefore, we are going to make a numerical analysis not only of the tree-level potential but also of the one-loop effective potential with and without the consideration of charge-breaking VEVs. For doing that, we use \Vevacious\cite{Camargo-Molina:2013qva} for which we have generated model files with \SARAH\cite{Staub:2008uz,Staub:2009bi,Staub:2010jh,Staub:2012pb,Staub:2013tta,Staub:2015kfa}. We also used \SARAH to generate a \SPheno module \cite{Porod:2003um,Porod:2011nf} for the NMSSM. With this module we calculate the SUSY and Higgs masses including NMSSM-specific two-loop corrections \cite{Goodsell:2014bna,Goodsell:2015ira,Goodsell:2016udb} which are important in particular for large $|\lambda|$ \cite{Goodsell:2014pla,Staub:2015aea}. Consequently, the accuracy in the Higgs mass prediction is similar to the MSSM and we use 3~GeV for the theoretical uncertainty in the following.
The spectrum file generated by \SPheno is passed to {\tt HiggsBounds} \cite{Bechtle:2008jh,Bechtle:2011sb} and is also used as input for \Vevacious. \Vevacious finds all solutions to the tree-level tadpole equations by using a homotopy continuation implemented in the code {\tt HOM4PS2} \cite{lee2008hom4ps}. These minima are used as the starting points to find the minima of the one-loop effective potential using {\tt minuit} \cite{James:1975dr}. If it finds deeper minima than the EW one, \Vevacious calls {\tt CosmoTransitions} \cite{Wainwright:2011kj} to get the tunnelling rate. However, in the standard \Vevacious package, the calculation for the tunnelling rate is not done for all minima, but only for the so called `panic' vacuum. This is the one closest to the EW minimum in field space. We have modified  \Vevacious to calculate the tunnelling rate to all minima in order to be able to compare the different sets of vacua.\\
We are going to distinguish two cases in the following: (i) cases in which only CB minima exist which are deeper than the EW one; (ii) cases in which both deeper CB and CC minima exist. In the following numerical examples, we will minimise the impact of the stop- and sbottom-sector on both Higgs mass and vacuum stability by assuming negligible trilinear couplings.
\subsection{Charge-breaking minima only}
\label{subsec:CBonly}
\begin{figure}[tb]
\centering
\includegraphics[width=0.5\textwidth]{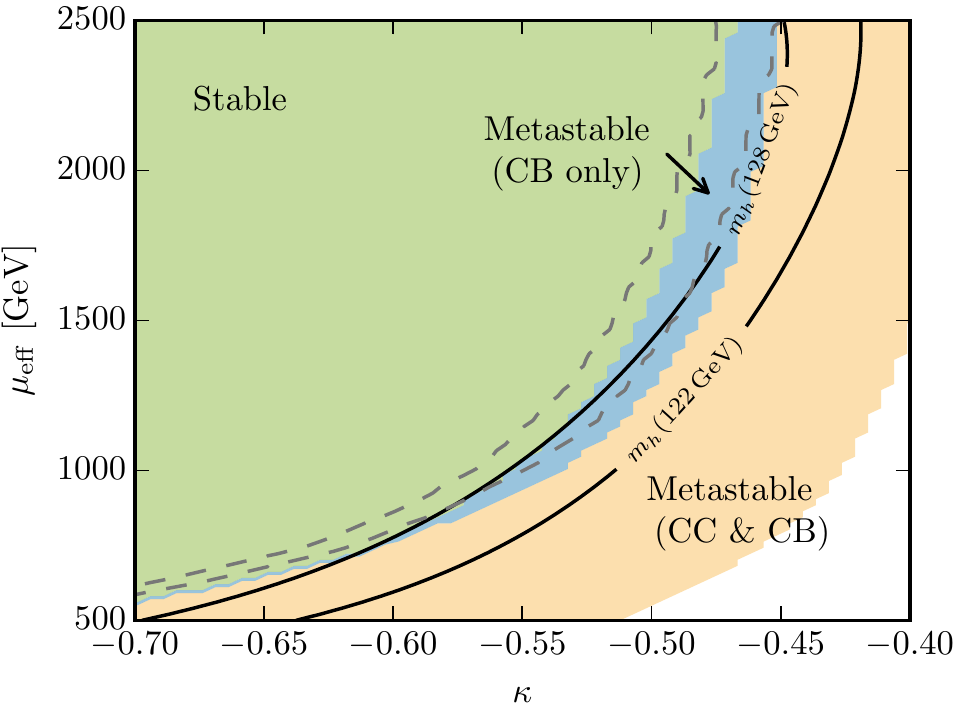} 
\caption{Stability of the EW vacuum considering the full one-loop effective potential. Regions shaded in green are stable, indicating that the desired electroweak breaking minimum is the global minimum. The yellow and blue regions correspond to metastablity of the desired electroweak breaking minimum. In particular, the blue region contains only CB minima that are deeper, while the yellow regions contains both CB and CC minima. The dashed-grey contours show the equivalent of the blue CB metastable region assuming only a tree-level potential. Finally, the region between the black solid contours corresponds to an acceptable Higgs mass, namely $m_h \in [122,128]\,\rm{GeV}$. Here we have chosen $\lambda=-0.68$, $\tan\beta =1.02$, $A_\kappa = -700\,\rm{GeV}$ and $A_\lambda=-300\,\rm{GeV}$. }
\label{fig:only_metastable_CB}
\end{figure}

Although it is not reflected in the analytical example discussed in Sec.~\ref{sec:analytic}, there also exist parameter points for which the EW minimum is only metastable once the possibility of charge breaking is included. Without the consideration of charged-Higgs-VEVs, the wrong impression of a stable EW minimum would be obtained. 
An example is shown in Fig.~\ref{fig:only_metastable_CB} where the blue region features a global CB minimum while the next-deepest minimum is the desired EW one. In the green region, the EW vacuum is stable whereas in the yellow region,  other CC minima corresponding to Eq.~(\ref{eq:dircon}) are also deeper than the desired EW one. In this figure, no parameter point which predicts the correct Higgs mass features a stable vacuum once the CB direction is taken into account. As a side remark we note that one can also see in this example that loop corrections to the scalar potential 
can be important when discussing the vacuum stability: if one would not have included charged Higgs VEVs, the conclusion whether stable regions in agreement the Higgs mass measurement exist would have changed from tree- to loop-level.\\

When checking all cases which we found in our scans, there were no points featuring  only CB minima  deeper than the desired EW one which turned out to be short-lived on cosmological time scales. All points had a life-time which was many orders of magnitude longer than the life-time of the universe. We therefore conclude that such points are phenomenologically viable, albeit significantly less appealing compared with regions where the vacuum is entirely stable.  \\

%
\subsection{Charge-breaking and charge-conserving minima}
\label{subsec:CBandCC}

\begin{figure}[tb]
\centering
\includegraphics[width=0.5\textwidth]{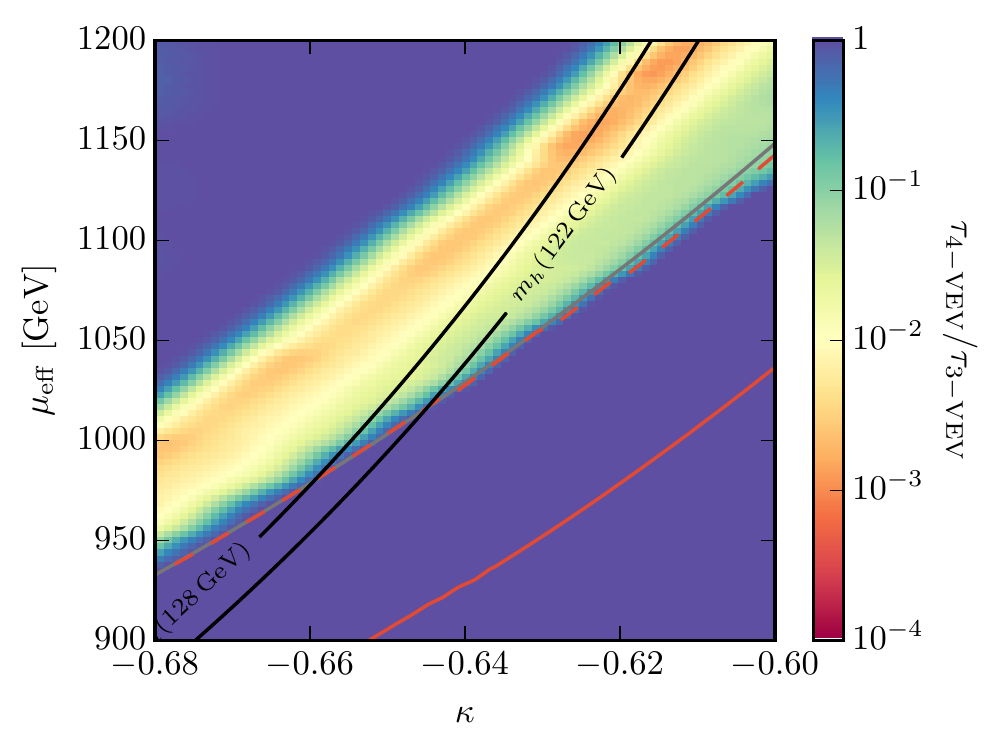} 
\caption{
Ratio of the lifetimes $\tau_{4\rm{-VEV}}$ and $\tau_{3\rm{-VEV}}$. Here, $\tau_{4\rm{-VEV}}$ and $\tau_{3\rm{-VEV}}$ are the lifetimes for the most unstable minima of the respective systems. The regions above the red (both solid and dashed) and grey contours correspond to at least 99\% survival probabilities of the DSB vacuum. The dashed red and solid grey contours correspond to the most unstable minima of the 3 and 4 VEV systems respectively. The solid red contour corresponds to the  stability of the panic vacuum (the minimum closest in field space to the DSB vacuum) in the 3 VEV system. Once again the the region between the black solid contours corresponds to an acceptable Higgs mass, namely $m_h \in [122,128]\,\rm{GeV}$. Here we have chosen $\lambda=-0.81$, $\tan\beta =1.02$, $A_\kappa = -1400\,\rm{GeV}$ and $A_\lambda=-580\,\rm{GeV}$.}
\label{fig:lifetime}
\end{figure}

This subsection aims to answer the question whether or not CB minima can further destabilise already metastable regions of parameter space, reducing the EW vacuum to be dangerously short-lived on cosmological time scales. As discussed before, this is not the case in regions where only CB minima are deeper than the EW minimum, which is why we turn to regions where also other CC minima are deeper. Indeed we find many regions of parameter space where the CB vacuum configuration corresponds to the global minimum,  with potential values up to $\mathcal O( 30\%)$ deeper compared to the next deepest CC minimum. However, as already 
seen in Fig.~\ref{fig:ana_comparison}, other non-EW CC vacua are nearer to the EW vacuum configuration in field space, which means that the tunnelling path is reduced compared to the tunnelling to the global, CB minimum. In practice, it turns out that this effect is more important than the relative depth of the minima. Although the global minimum is often CB, we find that the tunnelling-time to the slightly nearer shallower CC configuration of Eq.~(\ref{eq:dircon}) is either shorter or of comparable size in the regions where the lifetime of the vacuum is comparable to the lifetime of the Universe.\footnote{Note that one can not generalise the statement that tunnelling to the nearer minimum is more effective: if we were to always consider the nearest minimum to the EW one, we would often underestimate the actual tunnelling rates by several orders of magnitude, as is also reflected in the numerical example shown in Fig.~\ref{fig:lifetime}.}
Furthermore, we find that in those few cases where the tunnelling to the CB minimum indeed results in a shorter lifetime, the differences are typically small. 
This behaviour is shown Fig.~\ref{fig:lifetime}. The background colours depict the ratio of the lifetimes when considering both CC and CB minima (denoted as $\tau_{4-\rm VEV}$) versus when considering only CC minima ($\tau_{3-\rm VEV}$). 
Purple correspond to regions where the tunnelling rate of the EW vacuum is unchanged when also considering the charged-Higgs VEVs. Deviations from the purple background colour indicate that including the charged-Higgs direction leads to a more effective tunnelling than only considering the neutral Higgs directions.
Regions above and to the left of the red dashed and grey contour correspond to parameter space where the vacuum is sufficiently long-lived for the 3- and 4-VEV systems respectively. Here, we have used a survival probability of 99\% to determine these contours. \\
Note that we see a slight difference between those lines in the upper right part of the figure. This is where the tunnelling to the CB minimum is more efficient than the tunnelling to the CC one. The area which the two lines enclose is, however, very small. Therefore  the inclusion of the charged-Higgs direction in the vacuum stability consideration barely changes the conclusion. The red solid line depicts the instability bound we would arrive at if we considered only the panic vacuum, i.e. the minimum nearest to the EW one in field space. It is therefore evident that a na\"ive check for the vacuum stability can severely underestimate the excluded parameter ranges.

All in all, we find that although the global minimum of an NMSSM parameter point can feature a global minimum where the charged Higgs develops a VEV, it is not necessary to check for this extra field direction as the constraints on the model parameters remain approximately unchanged if one ensures that the tunnelling rate to all possible minima are calculated.

 \vskip1cm 

\section{Conclusion}
\label{sec:conclusion}
We considered the possibility of spontaneous charge breaking in the NMSSM via VEVs of the charged Higgs components. We found that in contrast to models without singlets it is possible that charge is broken at the global minimum of the potential. We could identify two different kinds of parameter regions. First, regions in which all vacua deeper than the electroweak minimum have broken electric charge. These points would give the wrong impression of a stable electroweak vacuum if charged Higgs VEVs were not included in the study. However, in all examples we found for these scenarios, the life-time of the electroweak vacuum is sufficiently long on cosmological time-scales. The second possibility is that charge-breaking and -conserving minima beside the EW one are present at the same time. Here, the charge-breaking minima could be significantly deeper than the charge-conserving ones. However, we found that the parameter regions which are excluded due to an increased tunnelling rate to these deeper vacuum states are hardly affected when considering the extra charged Higgs field direction. Thus, the inclusion of charge-breaking minima doesn't drastically change the conclusion of a `long-lived' vacuum to a `short-lived' one. All in all, despite the presence of deep charge-breaking minima in the NMSSM, their phenomenological impact  is rather limited. However, we want to stress that the usual practice of checking only the tunnelling rate to the  deeper minimum nearest to the electroweak vacuum is insufficient for obtaining reliable bounds on the NMSSM parameter space.

\section*{Acknowledgements}

M.E.K is supported by the DFG Research Unit 2239 ``New Physics at the LHC''. T.O is supported by the SFB-Transregio TR33 ``The Dark
Universe''.

\bibliographystyle{ArXiv}
\bibliography{lit}

\end{document}